\documentclass[a4paper,11pt]{article}
\usepackage{pos}
\usepackage{xcolor}
\usepackage{natbib}

\renewenvironment{thebibliography}[1]{
  \begin{oldthebibliography}{#1}
    \setlength{\itemsep}{0.5em}
    \setlength{\parskip}{0em}
}
{
  \end{oldthebibliography}
}

\setcitestyle{author,open={[},close={]}} 

\usepackage{booktabs, tabularx}

\title{Optical Microlensing by Primordial Black Holes with IACTs}
 \ShortTitle{Optical Microlensing by Primordial Black Holes with IACTs}

\author*[a]{Konstantin Pfrang}
\author[b]{Tarek Hassan}
\author[a]{Elisa Pueschel}

\affiliation[a]{Deutsches Elektronen Synchrotron DESY,\\
  Platanenallee 6, 15738 Zeuthen, Germany}

\affiliation[b]{Centro de Investigaciones Energéticas, Medioambientales y Tecnológicas (CIEMAT)\\
Av. Complutense, 40, E-28040 Madrid, Spain}

\emailAdd{konstantin.pfrang@desy.de}

\abstract{Primordial black holes (PBHs), hypothesized to be the result of density fluctuations during the early universe, are candidates for dark matter. When microlensing background stars, they cause a transient apparent enhancement of the flux. Measuring these signals with optical telescopes is a powerful method to constrain the PBH abundance in the range of $10^{-10}\,M_{\odot}$ to $10^{1}\,M_{\odot}$. Especially for galactic stars, the finiteness of the sources needs to be taken into account. 
For low PBH masses (in this work $\lesssim 10^{-8}\,M_{\odot}$) the average duration of the detectable event decreases with the mass $\langle t_e\rangle \propto  M_{\mathrm{PBH}}$.
For $M_{\mathrm{PBH}}\approx 10^{-11}\,M_{\odot}$ we find $\langle t_e\rangle \lesssim\,1 \mathrm{s}$.
For this reason, fast sampling detectors may be required as they could enable the detection of low mass PBHs.
Current limits are set with sampling speeds of 2 minutes to 24 hours in the optical regime. Ground-based Imaging Atmospheric Cherenkov telescopes (IACTs) are optimized to detect the $\sim$ns long optical Cherenkov signals induced by atmospheric air showers. 
As shown recently, the very-large mirror area of these instruments provides very high signal to noise ratio for fast optical transients ($\ll 1\,$s) such as asteroid occultations.
We investigate whether optical observations by IACTs can contribute to extending microlensing limits to the unconstrained mass range  $M_{\mathrm{PBH}}<10^{-10}M_\odot$. We discuss the limiting factors to perform these searches for each telescope type. We calculate the rate of expected detectable microlensing events in the relevant mass range for the current and next-generation IACTs considering realistic source parameters.}

\FullConference{37$^{\rm{th}}$ International Cosmic Ray Conference (ICRC 2021)\\
		July 12th -- 23rd, 2021\\
		Online -- Berlin, Germany}

\begin{document}
\maketitle

\section{Introduction}
\begin{figure}[htbp]
    \centering
    \includegraphics[width=0.65\textwidth]{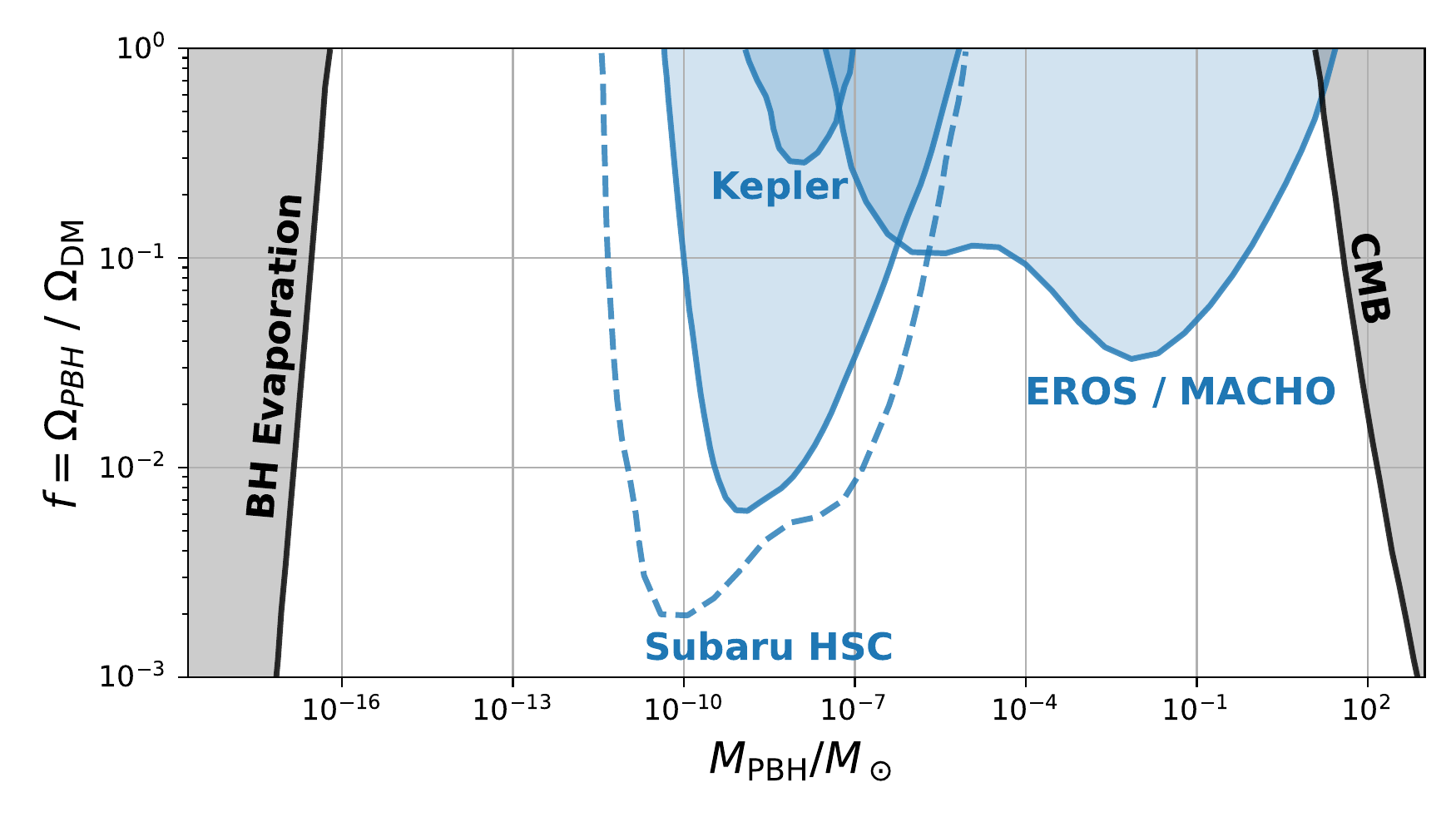}
    \caption{Overview of PBH abundance limits. Microlensing constraints are shown in blue.}
    \label{fig:overview}
\end{figure}

Dark matter (DM) is among the most fundamental ingredients for structure formation in the Universe \cite{structure}.
Although it is widely observed by its gravitational interaction, a direct detection remains elusive.
The nature of DM remains among the most important problems in physics.
Primordial black holes (PBHs) were first proposed in \cite{hawking_dm} as a possible DM candidate.
The various possible PBH formation mechanisms result either in a broad or narrow spectrum of PBH masses spanning almost twenty orders of magnitudes.
In attempts to constrain the PBH abundance over the full mass range, different experimental concepts are deployed.
An overview of current limits on the PBH abundance is shown in \autoref{fig:overview}.
Gravitational microlensing is a powerful method currently constraining over eleven orders of magnitude in PBH mass.
These limits are shown by the blue regions in \autoref{fig:overview}
During a microlensing event, a time-varying magnification of a background star can be observed when a compact object crosses the line of sight. 
By monitoring stars in the Large Magellanic Cloud with roughly 24-hour cadence, the MACHO and EROS experiments have constrained the PBH abundance in a mass range of $[10^{-7}, 10]\,M_\odot$ \cite{MACHO, EROS}. 
Observation of stars in the Galactic bulge by OGLE with 20 to 60 minutes sampling revealed a population of six short events, which could be well explained by either free-floating planets or PBHs \cite{OGLE}.
A microlensing study on Galactic stars was performed with a sampling speed of 30 minutes using 2 years of Kepler data that constrained PBH masses down to $10^{-8}\,M_\odot$ \cite{kepler_exp}.
Recently, Andromeda observations with the Subaru Hyper Suprime-Cam (HSC) were carried out to search for microlensing with a dense cadence of 2 minutes \cite{subaruHSC}. The sensitivity to short events made it possible to extend microlensing limits on PBH abundance to $10^{-11}\,M_\odot$, which is shown by the dashed blue line in \autoref{fig:overview}.
However, Ref. \cite{subaruupdate} pointed out that with an updated source size distribution the PBH constraints are limited to masses $>10^{-10}\,M_\odot$, shown by the solid blue line.
Microlensing PBH limits are adjacent to the currently unconstrained mass range of $[10^{-16}, 10^{-10}]\,M_\odot$.
At low and high masses the limits are enclosed by constrains from black hole evaporation and accretion effects on the CMB \cite{bh_evap, cmb}.

IACT arrays are designed to detect very-high-energy (VHE $E>100\,\mathrm{GeV}$) gamma-ray photons.
As a gamma ray interacts with the atmosphere, it triggers an atmospheric air shower. 
IACTs are optimized to detect the nanosecond-timescale Cherenkov flashes produced in the ultraviolet and blue bands within these showers.
For this, they are equipped with photodetectors sampling up to GHz frequencies. 
Beyond their abilities in gamma-ray astronomy, recent works proved that IACTs may also operate as competitive telescopes to detect very fast transient optical signals.
In \cite{asteroid_nature}, the high signal to noise achieved over millisecond timescales allowed the detection of the diffraction pattern produced during asteroid occultations. 
Via a diffraction fitting technique they were able to perform direct measurement of two stellar diameters with unprecedented resolution \cite{IACToptical}. 

In the following, we investigate the possibility to use IACTs to search for PBH-induced microlensing signals in the optical range. 
As an example of the current generation of IACTs we refer to VERITAS, located at the Fred Lawrence Whipple Observatory (FLWO) in southern Arizona (31 40N, 110 57W, 1.3km a.s.l.).


\section{Gravitational microlensing}\label{sec.micro}
Microlensing events are observed by monitoring the apparent brightness of stars.
A PBH located close to the line of sight (LoS) between a background star and Earth acts as a gravitational lens. 
In the case of microlensing, the multiple images that are created in this process are not individually resolvable and an apparent brightening of the star is observed. 
This leads to a time-dependent apparent amplification of the star magnitude.
Especially for nearby stars, the extension of the source has to be considered.
The amplification in the finite source case $A_{\mathrm{FS}}(U_*)$ is described by equations (9)-(11) of \cite{wittmao}.
The extension is encoded in the projected star radius $U_*$ that depends on the the radius of the star $R$, the ratio between the distance of the lens to the distance of the star $x$ and the Einstein radius $r_\mathrm{E}$.
The Einstein radius is the characteristic angle for gravitational lensing and depends on the the distances in the system as well as the PBH mass.
The projected radius increases with $U_* \propto M_{\mathrm{PBH}}^{-0.5}$ making it more crucial for low $M_\mathrm{PBH}$. 
An event is detectable during the time in which the amplification exceeds the threshold for detection $A_\mathrm{thresh}$ of the instrument.
Thus, the detectable event duration depends on the instrument and increases as the sensitivity of the instrument to flux changes improves.

In this work, we study the potential for IACTs to detect PBHs in the unconstrained mass range $<10^{-10}M_\odot$. 
The finite source amplification is limited to $A_{\mathrm{max}, \mathrm{FS}}$.
For PBHs that are closer to the source star, $x\rightarrow1$, the maximum amplification decreases.
This results in a point $x_{\mathrm{max}}$ that is the limit for the PBH distance at which the amplification still could be detected by the instrument. 
For heavier PBHs, the projected radius is small and $x_{\mathrm{max}}\approx1$. However, for lighter PBHs it decreases proportional to the PBH mass $x_{\mathrm{max}}\propto M_{\mathrm{PBH}}$.
The exact point of transition depends on the star parameters such as radius $R$, magnitude $m$, and distance $D_s$ as well as the sensitivity of the instrument to detect the transient signal $A_\mathrm{thresh}$.
For the current generation of IACTs and the sampling of $50\,\mathrm{Hz}$ selected in this study, this transitions is around $M_{\mathrm{PBH}}\approx 10^{-8}M_\odot$ for the best target. 
Following \cite{kepler2011}, the optical depth for the finite source limit can be calculated using:
\begin{align}\label{tau}
    \tau_{\mathrm{FS}} \approx \frac{\pi D_s\rho_0 R^2x_{\mathrm{max}}^3}{3 M_{\mathrm{PBH}}},
\end{align}
where $\rho_0\approx7.9\times10^{-3}\,M_\odot / \mathrm{pc}^3$ is the local dark matter density.
As will be discussed in \autoref{sec.IACTs}, close-by, bright stars are the best targets for IACTs. 
Thus, we can assume the constant local DM density along the LoS.
We use the approximate formula for the total event rate from \cite{kepler2011}
\begin{align}\label{gamma}
    \Gamma_{\mathrm{FS}} \approx 2 \frac{\tau_{\mathrm{FS}} v_c}{\pi x_{\mathrm{max}}R}.
\end{align}
where $v_c\approx220\,\mathrm{km\,/\,s}$ is the halo circular velocity.
For larger $M_\mathrm{PBH}$, where $x_\mathrm{max}\approx1$, the lower total number of PBHs in the DM halo leads to $\Gamma_{\mathrm{FS}} \propto M_\mathrm{PBH}^{-1}$.
However, due to the correlation of $x_\mathrm{max}$ to $M_\mathrm{PBH}$ below this transition, we find $\Gamma_{\mathrm{FS}} \propto M_\mathrm{PBH}$.
The duration in which the event might be detectable is given by the average event duration
\begin{align}
    \langle t_e\rangle = \frac{\tau_{\mathrm{FS}}}{\Gamma_{\mathrm{FS}}}.
\end{align}
This duration is directly correlated with $x_\mathrm{max}$ and thus, also a function $M_\mathrm{PBH}$.
In the range of $x_\mathrm{max}<1$, we find $\langle t_e\rangle\propto M_\mathrm{PBH}$.
Thus, the sensitivity to fast optical transients, could allow the detection of microlensing events with small $M_\mathrm{PBH}$.

\section{PBH Microlensing Observations with IACTs}\label{sec.IACTs}
The existing PBH studies use traditional optical instruments.
These provide a good optical precision combined with the simultaneous monitoring of up to $10^8$ stars \cite{subaruHSC}.
This is possible due to the use of CCD cameras with a high pixel density. 
On the other hand, these instruments are limited to a minimum of 2 minutes sampling speed.
Compared to these, IACTs are optical telescopes that are constructed to detect the $\sim\mathrm{nanosecond}$ flashes of Cherenkov light from atmospheric air showers.  
For an overview of IACTs as optical instruments, see \cite{Daniel}.
Their large optical reflectors make them powerful detectors for high time resolution photometry, as they minimize atmospheric scintillation noise.
Their overall optical precision however is modest, compared to traditional optical instruments. 
VERITAS is sensitive to measure the flux of objects with $\sim10.2$ magnitude at 2,400\,Hz with uncertainties of $10\%$ \cite{asteroid_nature}.
Each of the VERITAS cameras consists of 499 pixels made of photomultiplier tubes (PMTs) that cover the field of view of 3.5 degrees \cite{veritas_pixel}. 
As each pixel is integrating the background light of a large part of the sky, only a bright target in the foreground would be a feasible candidate for PBH searches.
In this study, we conservatively assume that only one bright foreground object possibly could be detected at once.

\subsection{Target Selection}
\begin{figure}
    \centering
    \includegraphics[width=\textwidth]{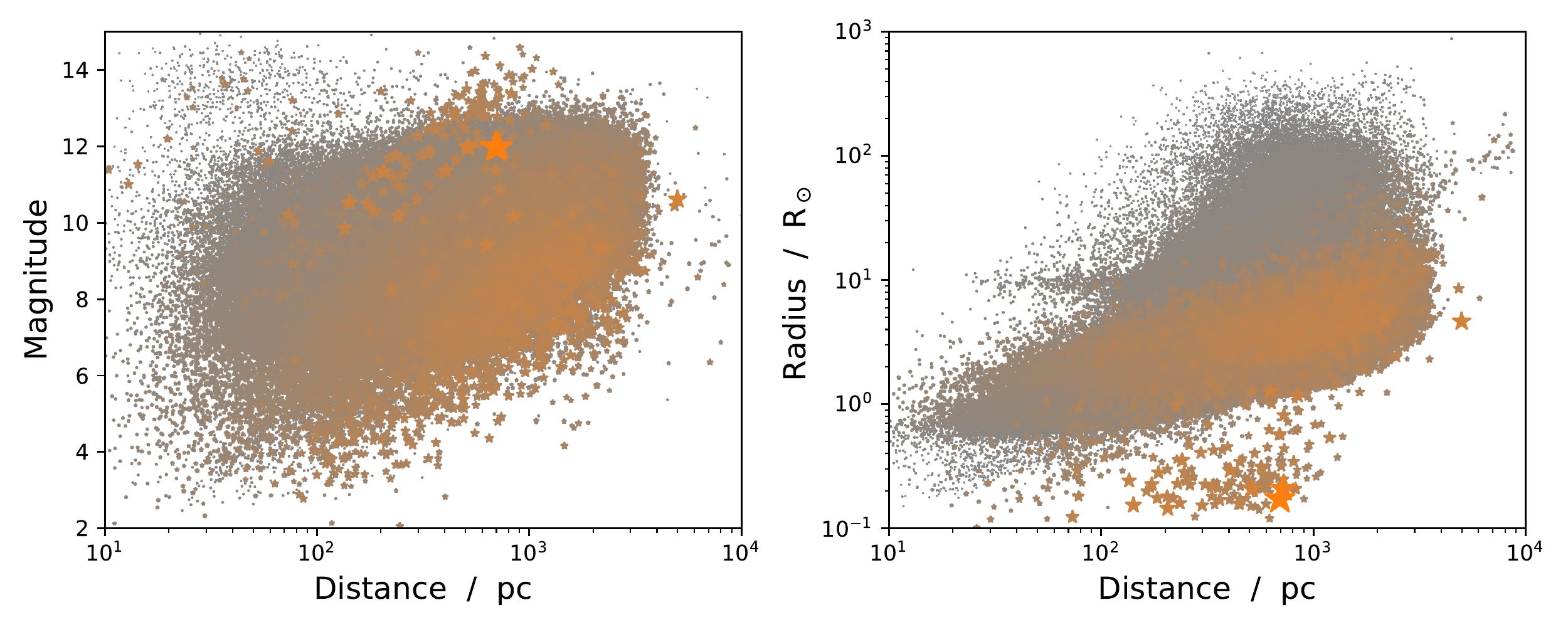}
    \caption{Distribution of relevant star parameters. 
    Large markers and orange color correspond to a high expected number of microlensing events in the mass range $M_\mathrm{PBH}<10^{-10}\,M_\odot$.}
    \label{fig:stars}
\end{figure}

In the following, we investigate the optimal target star to constrain the abundance of PBHs below $10^{-10}\,M_{\odot}$ with VERITAS.
We only consider shot noise for which the relative uncertainty decreases as $N^{-0.5}$ with the total number of photons $N$.
The total photon counts consist of the source $N_\mathrm{src}$ and the night sky background (NSB) $N_\mathrm{bck}$ contribution. 
We assume a constant NSB level of magnitude $m=9$ and scale the relative uncertainties on the measured star fluxes to arbitrary magnitudes.
The sensitivity to detect flux changes $A_\mathrm{thresh}$ is embedded into this uncertainty.
For bright stars this value is roughly constant $A_\mathrm{thresh}\sim1$.
At higher magnitudes $A_\mathrm{thresh}$ increases exponentially with $m$.
For the sensitivity of VERITAS with the given sampling speed in this study of 50\,Hz, we find the transition to be around $m\approx13$.
In the mass range we consider in this work ($M_\mathrm{PBH}<10^{-10}\,M_\odot$), this influences the maximum distance $x_\mathrm{max}$.
For $m\lesssim13$ we find $x_\mathrm{max}\propto 10^{-0.2m}$ and above $x_\mathrm{max}\propto 10^{-0.8m}$.
To describe the smooth transition we use a linear interpolation $f(m)$. 
With this, $x_\mathrm{max}$ can be described by the stellar parameters and the PBH mass:
\begin{align}\label{xmax}
    x_{\mathrm{max}} \propto \frac{D_s f(m)}{R^2}M_\mathrm{PBH}.
\end{align}
By inserting \autoref{xmax} and \autoref{tau} into \autoref{gamma}, the dependency of the total event counts in this regime can be written as
\begin{align}
    \Gamma_\mathrm{FS} \propto \frac{D_s^3 f(m)^2}{R^3} M_\mathrm{PBH}.
\end{align}
The ideal target star is selected to optimize the expected event rate.
It is a trade off between the distance of the star (corresponding to a large amount of DM along the LoS), the radius and the magnitude of the star (accounting for a large amplification in a microlensing event).
We investigate objects in the JSDC catalog \cite{jsdc} that contains the V magnitude as well as the expected angular diameter of 482,723 stars.
Furthermore, we query the SIMBAD database \cite{simbad} to obtain the distances for these objects, resulting in 433,378 candidates.
The distribution of the relevant star parameters is shown in \autoref{fig:stars}.
The colors and size of the markers represent how well-suited they are as a potential target for a microlensing PBH search.
We find, that among the best target stars the majority are B-type stars. 
Due to their very small radii especially hot subdwarf stars (sdO/Bs), which are situated in the extreme horizontal branch \cite{subd, ehb}, are among the best targets.
The optimal target is the sdO/B star PG 0240+046 with a V magnitude 11.98, distance $D_s=692\,\mathrm{pc}$ and radius $R=0.174\,\mathrm{R_\odot}$.
We note, that further effects should be considered for the selection of the optimal target.
Among these are the possible saturation of the pixels of the instrument for bright stars and the possible variability of the star.
The consideration of these is beyond the scope of this work.g
Furthermore, the JSDC catalog does only contain a small fraction of O-type stars that possibly also might be good targets due to the large ratio of brightness to radius.

\subsection{Event Rates}\label{sec.limits}
As discussed above, IACTs benefit from a very fast sampling speed compared to previous investigations. 
While in these cases, the low mass limits are often limited by the minimum event duration that could be detected, IACTs are less constrained here.
Therefore, we do not integrate the differential event rate $\frac{d\Gamma}{d t_e}$ over the relevant range of event duration (e.g., see \cite{griest1991, kepler2011, griest2013}).
Instead, the mainly limiting factor for IACTs is the low number of stars and the limited sensitivity to flux changes.
These are already considered in the total event rate $\Gamma$ given by \autoref{gamma}. 
We follow the typical assumption that the whole local dark matter consists of PBHs with a delta mass profile.
For simplicity, we assume that during the observations always one star with parameters as PG 0240+046 could be monitored, which is is an optimistic assumption.
Thus, the results should be considered as upper limits.
The counts $N$ develop linear with the sampling duration $t_s$ and the relative error due to shot noise decreases with $t_s^{-0.5}$. 
In this work, we assume that a $50\,$Hz sampling is used which results in a $\sim 1/7$ times lower relative uncertainty compared to the $2,400\,$Hz sampling of \cite{asteroid_nature}.
We require 4 consecutive samples to be enhanced by more than 3 sigma for the detection of a PBH event which corresponds to less than one fake positive per year.
\begin{figure}[htbp]
    \centering
    \includegraphics[width=\textwidth]{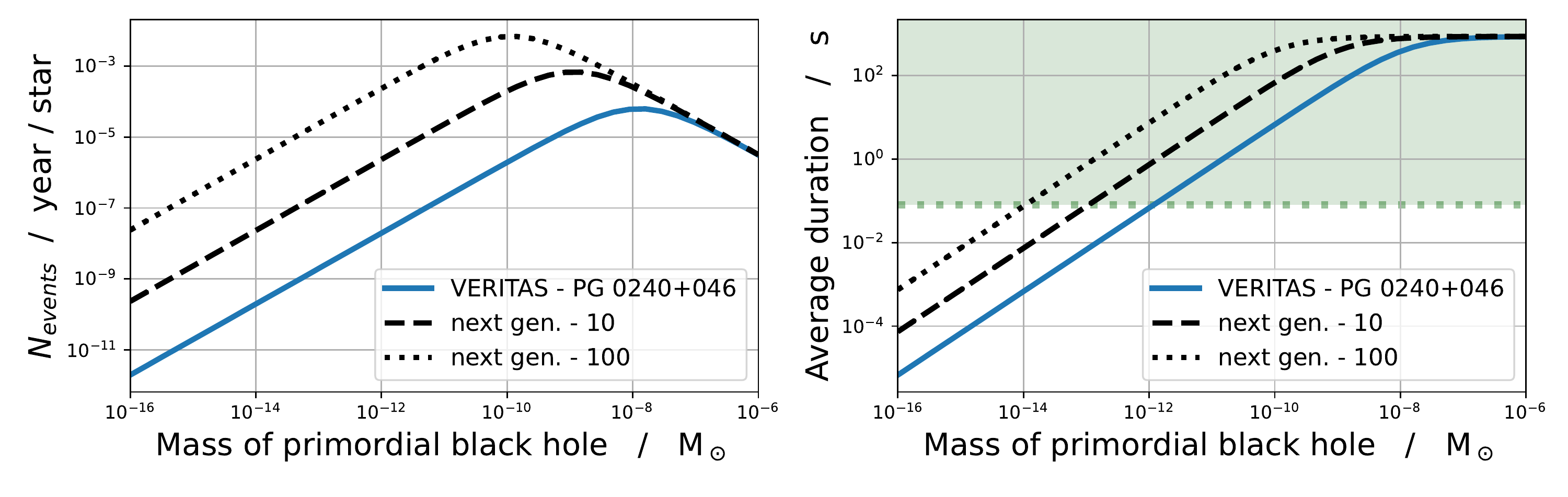}
    \caption{\textit{Left:} Expected event rates for the optimal target star PG 0240+046. The blue solid line is the expected event rate for VERITAS while the black dashed line a possible next-generation IACT array assuming 10 or 100 times improvements in sensitivity.
    \textit{Right:} Average duration of the expected events.}
    \label{fig:limit}
\end{figure}

The event rate with VERITAS as a function of the PBH mass are presented by the blue solid line in the left panel of \autoref{fig:limit}.
The transition of the $x_\mathrm{max}$ dependency is around $10^{-8}\,M_\odot$.
On the right graph, we show the average duration of these events. 
As expected, at the mass of interest, it decreases with $M_\mathrm{PBH}$ which is a consequence of the change in $x_\mathrm{max}$.
The green area shows the range, in which the average event duration would be detectable in 4 consecutive samples with $50\,Hz$.
With the given sampling speed, VERITAS could detect events down to $M_\mathrm{PBH}\approx 10^{-12}\,M_\odot$.
However, in the interesting mass range $M_\mathrm{PBH}<10^{-10}\,M_\odot$ such an event can only be expected every $\sim10^6\,\mathrm{years}$ of observation.

We also investigate the possible performance of a next-generation IACT.
The black dashed and dotted lines in \autoref{fig:limit} show the results assuming an improvement in the uncertainty of the flux measurement (and thus in $A_\mathrm{thresh} - 1$) by a factor of 10 and 100 respectively (accounting for e.g. larger mirror reflecting areas or photodetectors with better quantum efficiencies).
The event rates increase by a factor $(A_\mathrm{thresh} - 1)^{-2}$ and the event duration is larger by $(A_\mathrm{thresh} - 1)^{-1}$.
Nevertheless, with the 100 times improved sensitivity, in the relevant mass range an event can only be expected after $\sim 150$ years of observation time with the given sampling speed.

These results strongly depend on the sampling duration $t_s$. 
As described above, it influences the sensitivity $(A_\mathrm{thresh}-1) \propto 1/\sqrt{t_s}$ and thus at low PBH masses directly changes the event rate $\Gamma_\mathrm{FS} \propto t_s$ and duration $\langle t_e\rangle \propto \sqrt{t_s}$. 
E.g., observing with a 1\,Hz sampling, a next-generation instrument with 100 times improved sensitivity could expect $\sim0.05$ events per year of observation time at the peak $M_\mathrm{PBH}\approx10^{-11}\,M_\odot$ and the average duration would still be detectable.
Even with this tuning of the search, no large event rate is expected.
Additionally, IACTs are limited to observe during the night and a promising target star is not always expected in the FoV.

\section{Conclusions}\label{sec.conc}
Imaging Air Cherenkov telescopes have proven in the past, that they can be powerful instruments for fast optical astronomy.
In this work, we investigate the possibility of IACTs to detect microlensing of primordial black holes in the currently unconstrained mass range of $M_\mathrm{PBH}<10^{-10}\,M_\odot$.
At low PBH masses, the event duration correlates linearly with the PBH mass, making fast sampling speed with high signal to noise crucial.
We find that the limiting factors for the expected event rate are the modest optical accuracy and the small number of stars that IACTs could simultaneously observe.
No detectable PBH-induced microlensing events are expected over the whole VERITAS Observatory lifetime.
These searches are still not competitive even assuming an increase in the sensitivity by a factor of 100 for a next-generation instrument.
Besides the fast sampling speed, a good sensitivity to flux changes as well as the ability to monitor a large number of stars is also required to constrain the PBH abundance with masses at $M_\mathrm{PBH}<10^{-10}\,M_\odot$.

\section*{Acknowledgements}
K. Pfrang and E. Pueschel acknowledge the support of the Young Investigators Program of the Helmholtz Association. This research has made use of the SIMBAD database, operated at CDS, Strasbourg, France. This research has made use of the Jean-Marie Mariotti Center JSDC catalogue.

\bibliography{reference}

\end{document}